\documentclass{osa-article}

\journal{osajournal}
\articletype{Research Article}
\usepackage{mathrsfs}
\usepackage{multirow}
\usepackage{ulem}
\usepackage{appendix}
\newcommand{\dC}{$^{\rm o}$C}

\newcommand{\ms}{{m s$^{-1}$}}
\newcommand{\fp}{Fabry-P{\'e}rot}

\begin{document}

\title{Frequency stability of the mode spectrum of broad bandwidth {\fp} interferometers}

\author{Jeff Jennings,\authormark{1,2,*} Ryan Terrien,\authormark{1,3} Connor Fredrick,\authormark{1,2}  Michael Grisham,\authormark{4} Mark Notcutt,\authormark{4} Samuel Halverson,\authormark{5,6,7} Suvrath Mahadevan,\authormark{8,9} and Scott A. Diddams\authormark{1,2,**}}

\address{
\authormark{1}Time and Frequency Division, National Institute of Standards and Technology, Boulder, CO 80305, USA\\
\authormark{2}Department of Physics, University of Colorado, Boulder, CO 80309, USA\\
\authormark{3}Department of Physics and Astronomy, Carleton College, Northfield, MN 55057, USA\\
\authormark{4}Stable Laser Systems, Boulder, CO 80301, USA \\
\authormark{5}Jet Propulsion Laboratory, Pasadena, CA 91109, USA\\
\authormark{6}Massachusetts Institute of Technology, Cambridge, MA 02139, USA\\
\authormark{7}NASA Sagan Fellow\\
\authormark{8}Department of Astronomy \& Astrophysics, Penn State, University Park, PA 16802, USA\\
\authormark{9}Center for Exoplanets \& Habitable Worlds, University Park, PA 16802, USA\\
}

\email{\authormark{*}jeffrey.m.jennings@colorado.edu}
\email{\authormark{**}scott.diddams@nist.gov} %% email address is required


\begin{abstract}
When illuminated by a white light source, the discrete resonances of a {\fp} interferometer (FP) provide a broad bandwidth, comb-like spectrum useful for frequency calibration. We report on the design, construction and laboratory characterization of two planar, passively stabilized, low finesse ($\approx 40$) FPs spanning 380 nm to 930 nm and 780 nm to 1300 nm, with nominal free spectral ranges of 20 GHz and 30 GHz respectively. These instruments are intended to calibrate astronomical spectrographs in radial velocity searches for extrasolar planets. By tracking the frequency drift of three widely-separated resonances in each FP we measure fractional frequency drift rates as low as $1\times 10^{-10}$ day$^{-1}$. However we find that the fractional drift rate varies across the three sample wavelengths, such that the drift of two given resonance modes disagrees with the ratio of their mode numbers. We explore possible causes of this behavior, as well as quantify the temperature and optical power sensitivity of the FPs. Our results demonstrate the advancement of {\fp} interferometers as robust and frequency-stable calibrators for astronomical and other broad bandwidth spectroscopy applications, but also highlight the need for chromatic characterization of these systems.
\end{abstract}

\section{Introduction}
The {\fp} interferometer\cite{1901ApJ....13..265F} is one of the most basic and important optical instruments, with widespread use in both technical and fundamental applications ranging from telecommunications to gravitational wave detection\cite{1989fpih.book.....V}. Many of its earliest uses were in astronomical spectroscopy; analyzing the interferometer's fringe pattern while observing the Orion nebula allowed Fabry and Buisson to measure gas compositions and radial velocities \cite{1914ApJ....40..241B}. Extending this rich history, the {\fp} interferometer continues to play a critical role in astronomical spectroscopy today.

In this work we investigate vacuum gap {\fp} interferometers (here referred to as FPs; etalons; cavities) constructed with plane-parallel mirrors to be used as wavelength references for the calibration of high precision astronomical spectrographs\cite{Wildi2012,McCracken14,Halverson_2014,Reiners14,2015A&A...581A.117B}. When illuminated by a broadband light source the FP functions as a spectral filter and provides a dense, broad bandwidth, comb-like spectrum of lines spaced in frequency by the cavity free spectral range. For the astronomical spectroscopy we envision, the output of the FP is collected in an optical fiber that feeds the input of a spectrograph. The spectrograph also receives stellar light in a second fiber, and in its focal plane, the FP transmission spectrum then provides a calibration grid against which wavelength shifts in the stellar spectrum are measured. 

The FPs studied here are intended to serve as calibrators in exoplanet detection surveys that employ the Doppler radial velocity (RV) technique\cite{2010exop.book...27L}, wherein an unseen orbiting planet induces a periodic Doppler shift of the spectrum of the host star. The most important quality of the FP in this context is its dimensional stability, which is equivalent to the stability of its frequency modes, and is tied to its optical and mechanical properties. For the most demanding RV spectroscopy aimed at detecting a terrestrial mass extrasolar planet (exoplanet) in the habitable zone of the host star, the required RV precision is in the range of $\approx$0.1 {\ms} to 0.3 {\ms}, corresponding to a fractional calibration stability $\lesssim 3\times 10^{-9}$ over timescales of years.  In spite of recent progress with even the best cryogenic optical cavities \cite{2014OptL...39.5102H,Robinson2019} this level of stability remains uncertain over months to years, though it is now achievable with laser frequency combs (LFCs) \cite{2007MNRAS.380..839M, 2008EPJD...48...57B, 2008Natur.452..610L, 2008Sci...321.1335S, 2012OExpr..20.6631Y, 2012Natur.485..611W}. However passive FPs do remain a valuable tool for spectrograph calibration because of their broad spectral coverage ($\geq 1$ octave) and high stability ($\sim 10^{-10}$ fractional) over a period of days, as well as their simplicity and cost relative to a LFC. In this role the passive FP offers a secondary, independent calibration system provided its slow (ideally predictable) drift can be characterized with other frequency or wavelength references on timescales of weeks to months.

For robust use as a wavelength reference, the frequency stability of the FP transmission spectrum must be quantified and its sensitivities to environmental and experimental parameters understood. Here we describe laboratory measurements to characterize the frequency stability and drift of two FP calibrators of similar design and construction (low coefficient of thermal expansion materials, vacuum gap interferometric cavities, plane-parallel mirrors), but spanning different wavelength ranges as determined by the reflectivity response of their mirror coatings.  The first is a near-infrared (780 nm to 1300 nm) FP calibrator designed for the Habitable Zone Planet Finder (HPF) \cite{2014SPIE.9147E..7ZH} spectrograph at the 10 m Hobby-Eberly Telescope at McDonald Observatory.  The second is a visible (380 nm to 930 nm) FP calibrator for use with the NEID \cite{Schwab2016} spectrograph at the 3.5 m WIYN Telescope at Kitt Peak National Observatory. Using an auxiliary LFC and a suite of scanning continuous wave lasers, we precisely and accurately monitor the frequency of multiple FP resonances over periods as long as 1 month. This allows us to characterize the stability of the FP over timescales relevant for astronomical applications. In both FP systems we observe fractional drift rates that are predominantly linear at the level of $\approx 5\times 10^{-10}$ day$^{-1}$.

Our most significant observation is a wavelength-dependent drift of the FP modes that deviates from the prediction that $\Delta \omega/\omega$ be constant. This suggests that {\it knowledge of the frequency and drift rate of a single resonance of a planar FP may not be sufficient to characterize the relative drifts of all other modes}, as has been proposed for FPs serving as astronomical calibrators\cite{1538-3873-127-955-880, Sturmer2017}. We find that drift rates vary from 50 kHz day$^{-1}$ to 800 kHz day$^{-1}$ ($10^{-10} - 10^{-9}$ fractional; $0.1$ {\ms} to 1.0 {\ms} RV equivalent) across wavelengths spanning hundreds of nm. We suspect this behavior can be traced to sensitivities of the angular illumination, parallelism and flatness of the FPs; however the exact cause is not yet identified. 

In the following we present the instrument designs (Section~\ref{sec:design}) and measurement techniques (Section~\ref{sec:lab_charac_scheme}) we employ, results of the FP stability measurements (Section~\ref{sec:results}), and discussion of the design and environmental factors that influence this stability. Our results demonstrate the high stability performance of simple, plane-parallel FP etalons, while emphasizing the importance of additional characterization and mode structure modeling of these instruments to further progress their use as robust spectrograph calibrators.

\section{Background}
\label{sec:background}
The resonance condition for a FP requires that the round trip optical phase shift be an integer multiple of $2\pi$. This includes both the propagation phase $k2L$ and the reflective mirror phase shift $2\phi_r$ such that

\begin{equation} 
   2\pi m  = k2L + 2\phi_r,
 \end{equation}
where $m$ is an integer, $L$ the cavity length, $c$ the speed of light, and $k=\omega n/c$ the frequency-dependent propagation constant. We set the index of refraction $n=1$ in vacuum. This relationship implies that the discrete resonant frequencies $\omega_m$ (the cavity longitudinal modes) are 
\begin{equation} 
   \omega_m  = \frac{c} {L}(\pi m -\phi_r),
 \label{eqn:omega}
 \end{equation}
where $\phi_r$ is generally frequency dependent.

For an ideal, dispersionless FP cavity the frequency response of the resonance modes to changes in the cavity length $\Delta L$ can be illustrated by a simple \lq{}rubber band\rq{} model in which higher frequency modes undergo larger frequency shifts, but for which the fractional frequency shift $\Delta\omega_m / {\omega_m} = -\Delta L / L $ across the spectrum is constant. 
The number of modes between two resonance frequencies predicts their relative frequency response to a changing cavity length, analogous to equidistant marks on a rubber band whose separations disperse nonuniformly as the band is stretched. Within this simplified model, the ratio of the frequencies of two modes is independent of $\Delta L$ and given simply by the ratio of their mode numbers $\omega_m/\omega_l=m/l$. Similarly, in response to a cavity length change the fractional shift in frequency $\Delta \omega_m$ is equal to the ratio of mode numbers $\Delta \omega_m / \Delta \omega_l=m/l$.

A more realistic model of the FP cavity must also account for dispersion of the cavity medium and mirror coatings, as well as wavelength-dependent phase shifts arising from imperfections in the surface figure of the mirrors and their parallelism. 
In addition, cavity length changes occur on short (minute to hour) timescales in response to thermal fluctuations and mechanical stresses, as well as long (day to year) timescales due to a gradual aging of the etalon spacer material and slow thermal and mechanical changes in the environment. In FP cavities built from ULE glass and Zerodur that are intended for laser stabilization, the latter behavior has been seen to result in an exponential change of the cavity resonance frequencies that asymptotically approaches a constant. This \lq{}drift\rq{} rate is typically 1 kHz day$^{-1}$ to 100 kHz day$^{-1}$ over years \cite{Zhu92,Dubé2009,Stoehr:06} and on timescales of days -- weeks can be well approximated as linear. 
While resonance mode frequency response to various perturbations has been studied previously, to our knowledge a comprehensive model to account for the sum of their behaviors does not yet exist. We expand on previous studies by experimentally identifying, over broad bandwidths and in two independent cavities, the deviation of cavity mode frequencies from the ideal case of $\Delta \omega_m / \Delta \omega_l=m/l$.

\section{Experimental setup}
\label{sec:setup}

\subsection{Etalon design}
\label{sec:design}
The principle design requirements for etalons as RV spectrograph calibrators are broad bandwidth ($\approx 1$ octave), free spectral range (FSR=$c/(2L)$) of order 10 GHz, moderate finesse ($\sim 10$ to 100), and excellent passive short-term frequency stability. The broad bandwidth is needed to provide calibration across the full spectral range of modern echelle spectrographs, while the FSR is chosen to provide the maximum spectral calibration information. This is typically 3 to 5 times the spectrograph's resolution, e.g., for a spectrograph with resolution $R = \lambda / \Delta \lambda \approx 50\ 000$ centered at $\lambda = 1 \mu$m the preferred FSR is $\approx 20$ GHz. With the FSR fixed, the etalon's finesse then determines the linewidth $\Delta \nu$, with a preference for $\Delta \nu < R$ to allow measurement of the spectrograph's point spread function. This is balanced by the requirement that $\Delta \nu$ be sufficiently large for the etalon to efficiently pass the largest photon flux from a white light illuminator, and together these conditions set an etalon finesse $\mathscr{F} \approx 5 - 100$. This is much lower than typical values in etalons used for atomic spectroscopy and frequency metrology, where $10^5 \lesssim \mathscr{F} \lesssim 5 \times 10^5$ \cite{Salomon:88,2017PhRvL.118z3202M}.  However the desired short-term stability and drift is comparable to the state of the art, representing a relatively unexplored region of design parameter space. While recent technological demonstrations and in situ (at the telescope, \lq{}on-sky\rq{}) measurements with low finesse FPs have made evident the improved precision capabilities they can offer for RV spectrographs, there have been few published studies (though see \cite{2015A&A...581A.117B, 2017A&A...601A.102C, 2018SPIE10702E..76S, 2019A&A...624A.122C}) quantifying the performance of these FP calibrators in detail and verifying theoretically predicted behavior. 

\begin{table*}
\caption{Targeted and measured mechanical and optical properties of the near infrared and visible etalons under investigation. The NIR etalon is designed for the Habitable Zone Planet Finder (HPF) spectrograph, the visible etalon for the NN-EXPLORE Exoplanet Investigations with Doppler Spectroscopy (NEID) spectrograph.}
\begin{center}
\begin{tabular}{lcc}
\hline
& \bf{NIR etalon} (HPF) & \bf{Visible etalon} (NEID)  \\
\hline
Nominal FSR [GHz] & 30 & 20  \\ 
Etalon gap [mm] & 5.0 & 7.5  \\
Coating diameter [mm] & 9.0 & 20.0 \\
Spacer inner diameter [mm] & 15.0 & 40.0  \\
Spacer \&  mirror outer diameter [mm] & 25.4 & 50.8  \\
Reflectivity bandwidth [nm] & $780-1300$ & $380-930$ \\ 
Target finesse & $40(9)$ & $40(15)$ \\
\multirow{3}{7em}{Measured finesse} & 780 nm: 33 & 780 nm: 39 \\ 
& 1064 nm: 43 & 532 nm: 34 \\ 
& 1319 nm: 41 & 660 nm: 37 \\ 
Parallelism [$\mu$rad] & $<0.5$ & $<0.5$ \\
Mirror wedge [mrad] & 8.7 & 8.7 \\
\hline
\end{tabular}
\end{center}
\label{tab:etalons}
\end{table*}

Based on these criteria for an RV spectrograph calibrator, we have designed, built and characterized two FP etalons whose mechanical and optical parameters are given in {\bf Table~\ref{tab:etalons}}. {\bf Fig.~\ref{fig:setup}} shows the etalons' schematic and housing, as well as our characterization approach. Both etalons are fabricated by Light Machinery from ultra-low expansion (ULE) glass with specified zero crossing of the coefficient of thermal expansion near 30 {\dC}. There is a center bore along the optical axis and a small side vent hole. The etalon mirrors, also made from ULE glass, are wedged and have broad bandwidth anti-reflective (AR) coatings on the outer surfaces to minimize parasitic residual etalons. The mirrors are optically contacted to the spacer with the wedges aligned such that the outer surfaces are also parallel to reduce additional prismatic effects over the large bandwidth of both cavities. The larger dimensions of the visible etalon (relative to the NIR etalon) were chosen to improve alignment tolerances.

The housings for both etalons are similar in construction, and so we discuss the general features, while noting important differences here and in Table~\ref{tab:etalons}.  The etalons are held in temperature stabilized and evacuated aluminum chambers of dimensions $\approx 40$ cm $\times 40$ cm $\times 20$ cm.  Fiber feedthroughs are used to transmit light into the vacuum, and the broad bandwidth light is coupled into and out of the etalons with commercially available off-axis parabolic mirrors (OAPs).  The NIR etalon uses 780HP FP/APC single mode fibers, while the visible etalon employs endlessly single mode fibers (ESMFs; NKT Photonics LMA-5) feeding the OAPs. 

This inner stage is surrounded by thermal shielding in the evacuated chamber with an ion pump and commercial bench-top pump controller holding vacuum pressure $\leq 2 \times 10^{-7}$ Torr. A commercial bench-top temperature controller is used in conjunction with a thermoelectric heater (TEC) to hold the base of the thermal shielding at a nominal temperature of $32.5\ {\rm^oC}$ for the NIR etalon and $34.3 \ {\rm^oC}$ for the visible etalon. Independent thermistors positioned around the thermal shield are used to verify that the temperature is stable to better than $5 \times 10^{-3}\ {\rm^oC}$. Lab tests indicate the setpoint is within 0.5 {\dC} of the coefficient of thermal expansion (CTE) zero-crossing for the visible etalon but $\lesssim 5$ {\dC} from the zero-crossing for the NIR etalon. Fiber-to-fiber throughput at sampled wavelengths of 780 nm, 1064 nm and 1319 nm in the NIR etalon (780 nm, 532 nm and 660 nm in the visible etalon) is $>50\%$. A cutaway mechanical drawing indicative of both instruments is shown in Fig.~\ref{fig:setup}(a). 

\subsection{Laboratory characterization}
\label{sec:lab_charac_scheme}
\begin{figure}
\includegraphics[width=\columnwidth]{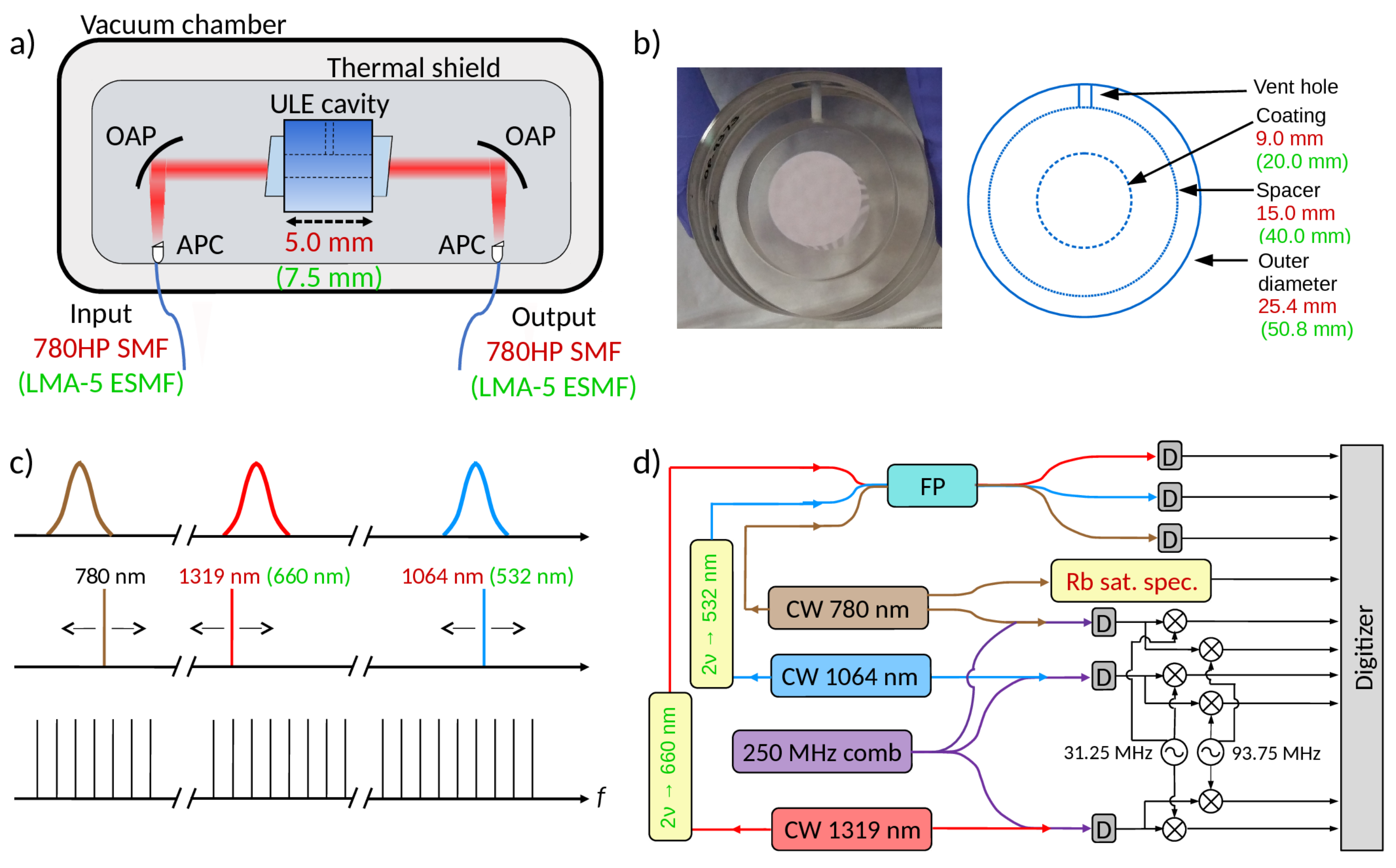}
\caption{NIR and visible etalons. a) Schematic of both {\fp} etalons under study. Cavity length and fiber types for the NIR etalon are in red text, those for the visible etalon in green. ULE, ultra low expansion glass; OAP, off-axis parabolic mirror; APC, angled physical contact connector; SMF, single mode fiber; LMA, large mode area; 780HP SMF, Thorlabs single mode fiber; LMA-5 ESMF, NKT Photonics endlessly single mode fiber. b) Photo of the visible etalon face-on and diagram showing components and scale (dimensions of NIR etalon in red, of visible etalon in green). c) Concept of the laboratory characterization scheme. CW lasers are scanned across spectrally displaced FP resonances while their frequency positions are tracked relative to a 250 MHz, self-referenced optical frequency comb. Frequencies in red text correspond to the NIR etalon setup, those in green to the visible etalon setup (780 nm is common to both systems). d) More detailed experimental configuration for both etalons, showing major optical and radio frequency components used to scan the FP resonances. Components in red text are unique to the NIR etalon setup (a portion of the 780 nm light in the NIR setup is sent through a rubidium saturated absorption cell), those in green text to the visible etalon setup (light from the 1064 and 1319 nm CWs is frequency doubled before entering the visible etalon). Three lasers are employed, with either the 780 nm or 1064 nm laser selected for a specific measurement in the NIR setup, while all 3 scan simultaneously in the visible setup. D, photodiode; Rb sat. spec., a rubidium saturated absorption spectroscopy setup; $2\nu$, frequency doubler.}
\label{fig:setup}
\end{figure}

To characterize the frequency stability of the FP cavities we monitor the frequencies of multiple cavity resonances relative to a self-referenced optical frequency comb. 
Full details of this technique are presented in \cite{2017OExpr..2515599J}; here we give a brief summary while noting differences from our previous work. A near infrared, self-referenced, octave spanning LFC \cite{Ycas2012} calibrates the scanning of continuous wave (CW) lasers at 780 nm, 1064 nm and 1319 nm with the NIR etalon (780 nm, 532 nm and 660 nm with the visible etalon, the latter two being the 1064 and 1319 nm laser light frequency doubled in nonlinear crystals) that are transmitted through the FP to simultaneously measure and track the frequencies of the resonance modes in real time. The CW laser power incident on both etalons at all wavelengths is $< 1$ mW.  Fig.~\ref{fig:setup}(c)--(d) give a schematic overview for both setups. The CW lasers are periodically scanned across their respective FP resonances (of width $\approx 500$ MHz to 900 MHz) by applying a sawtooth frequency scan with 40 s period for the NIR etalon, 100 s for the visible etalon.
 After transmission the beams are de-multiplexed in fiber and sent to separate detectors where the transmitted power from each laser is photodetected and digitized. In addition, our NIR etalon setup includes a portion of the 780 nm laser light being sent through a vapor cell containing rubidium (Rb), and saturated absorption of the Rb $\rm D_2$ lines are observed. These spectroscopic signals provide a secondary frequency reference for the 780 nm light to verify our primary frequency axis calibration technique (see below).

In separate detector channels the CW lasers are heterodyned with the spectrally broadened output of a self-referenced Er:fiber LFC with repetition rate $f_{\rm rep}=250$ MHz. In both 
setups a portion of the LFC near 1560 nm is frequency doubled in a nonlinear crystal to perform this heterodyne with the 780 nm laser. The LFC is stabilized to an in-house hydrogen maser, which is part of the NIST timescale. This ensures absolute optical uncertainty $<100$ Hz for all times $> 100$ s. 

As described in greater detail in \cite{2017OExpr..2515599J}, our electronic mixing scheme provides a calibration \lq{}tick\rq{} each time the CW laser moves 62.5 MHz, or at precisely $f_{\rm rep} / 4$. The known frequency spacing of the calibration ticks gives discrete points that map acquisition time to relative frequency. We fit a tenth order polynomial to this data to interpolate the frequency axis between calibration ticks and remove nonlinearities in the laser frequency scan.\footnote{One may expect the nonlinearity in the laser scans to induce a signal noise (\lq{}ringing\rq{}) in the calibration tick signal as the heterodyne beat between the CW laser and comb passes through the RF bandpass windows used to generate the ticks. We find no such discernible ringing and thus are confident that our Gaussian fits to individual ticks are not systematically biasing the separation between ticks over the nonlinear scan, i.e., that we are justified in assuming the separation between the fitted Gaussians to neighboring ticks is 62.5 MHz.} 

This calibrates the frequency axis for a single scan of the 780 nm laser to $<1.5$ MHz and the 1064 and 1319 nm lasers to $<100$ kHz in the NIR etalon setup. The larger uncertainty in the 780 nm signal is a consequence of the increased sensitivity of the distributed feedback laser source to noise on the sweep voltage controlling the laser current (contrasted with the 1064 nm and 1319 nm nonplanar ring oscillator lasers swept by thermal tuning). In the visible etalon setup the calibration uncertainty at 780 nm is the same, at 532 nm is $<300$ kHz and at 660 nm is $<1$ MHz. The higher uncertainty in the latter two relative to the NIR setup is due to scanning the lasers over a larger range, doubling the frequency axis and thus doubling its error to calibrate the frequency doubled 1064 nm and 1319 nm light sent into the visible etalon, and lower calibration tick SNR in the visible etalon setup.

\begin{figure}
\includegraphics[width=\columnwidth]{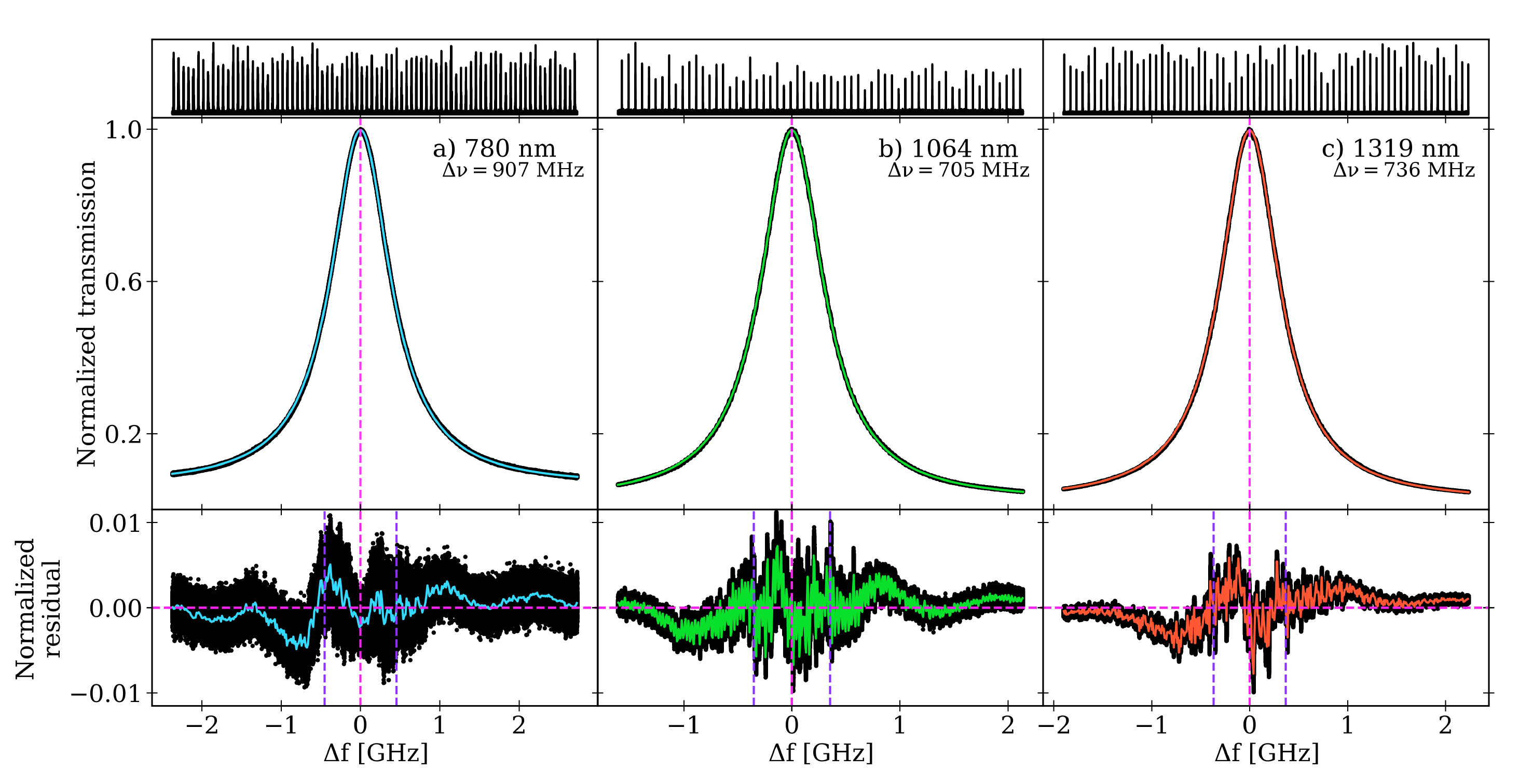}
\caption{NIR etalon. a) Normalized transmission of the etalon resonance at 384.2 THz ($\approx 780$ nm) for a single scan, with Lorentzian fit shown and fit linewidth $\Delta \nu$ given. The calibration \lq{}ticks\rq{} used to define and interpolate the frequency axis are shown above. Residuals to the Lorentzian fit at the $1\%$ level (normalized to the fit peak) are shown below with a 5 MHz rolling average and vertical lines indicating the fit peak and FWHM. b -- c) As in (a) for the resonances at 281.6 THz and 227.2 THz ($\approx 1064$ nm and 1319 nm).}
\label{fig:resonances}
\end{figure}

\begin{figure}
\includegraphics[width=\columnwidth]{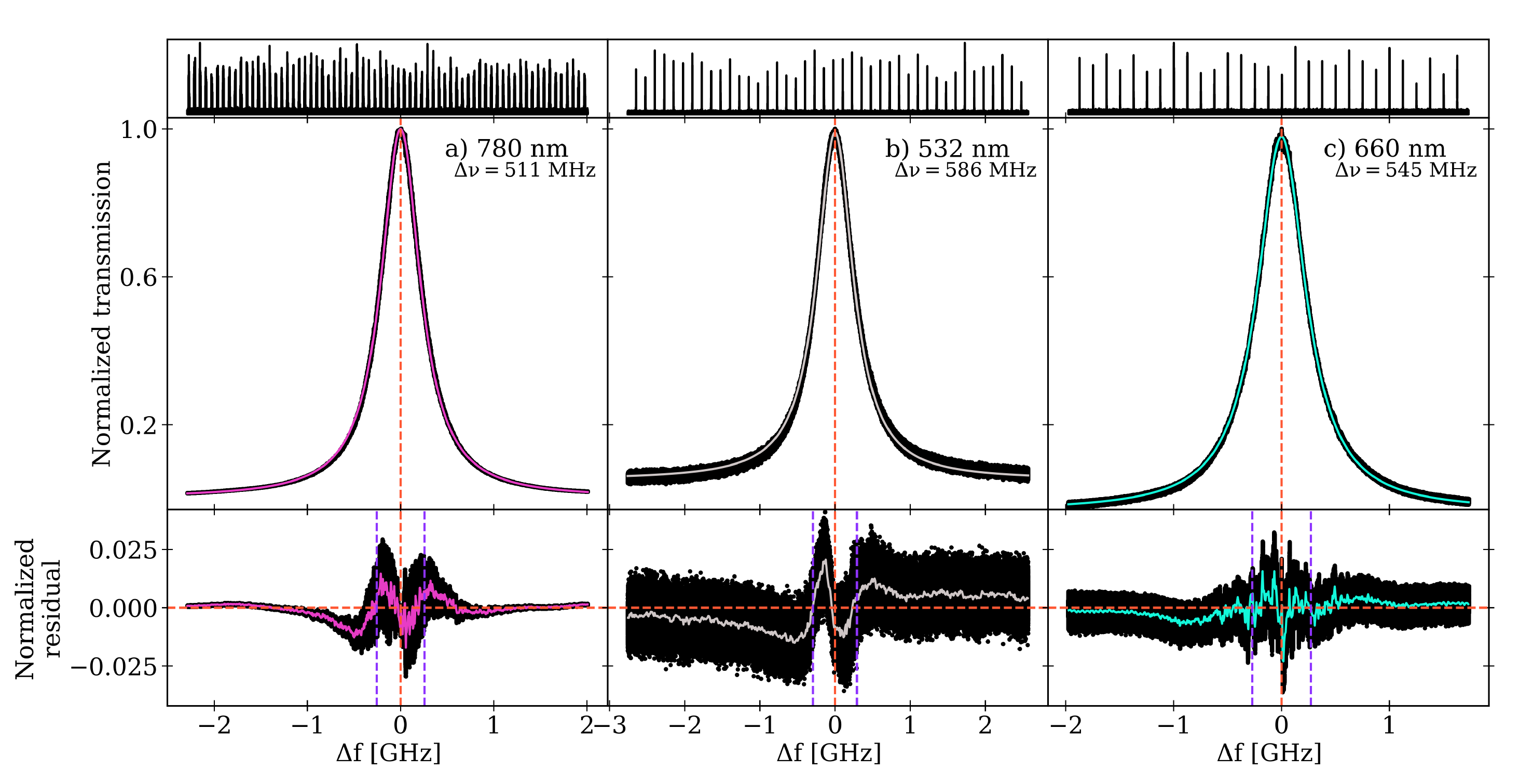}
\caption{Visible etalon. a) Normalized transmission of the etalon resonance at 384.2 THz ($\approx 780$ nm) for a single scan, with Lorentzian fit shown and fit linewidth $\Delta \nu$ given. The calibration \lq{}ticks\rq{} used to define and interpolate the frequency axis are shown above. Residuals to the Lorentzian fit at the few \% level (normalized to the fit peak) are shown below with a 5 MHz rolling average and vertical lines indicating the fit peak and FWHM. b -- c) As in (a) for the resonances at 563.2 THz and 454.4 THz ($\approx 532$ nm and 660 nm).}
\label{fig:resonances_visible}
\end{figure}

Using the respective frequency axis for each wavelength, the FP transmission resonance data are fit with a Lorentzian profile whose centroid is assigned as the peak resonance frequency. Example acquisitions of both the calibration tick signal and FP resonance scan at each of the three wavelengths in the NIR etalon are shown in {\bf Fig.~\ref{fig:resonances}} ({\bf Fig.~\ref{fig:resonances_visible}} in the visible etalon). The random (white noise) component of the residuals to the Lorentzian fits suggest an achievable noise-limited centroid measurement precision at or better than $\approx 2$ MHz in a single scan. However Fig.~\ref{fig:residual_struc} also demonstrates stable structure of unknown origin in the Lorentzian fit residuals, discussed in Appendix A. 

\begin{figure}
\includegraphics[width=\columnwidth]{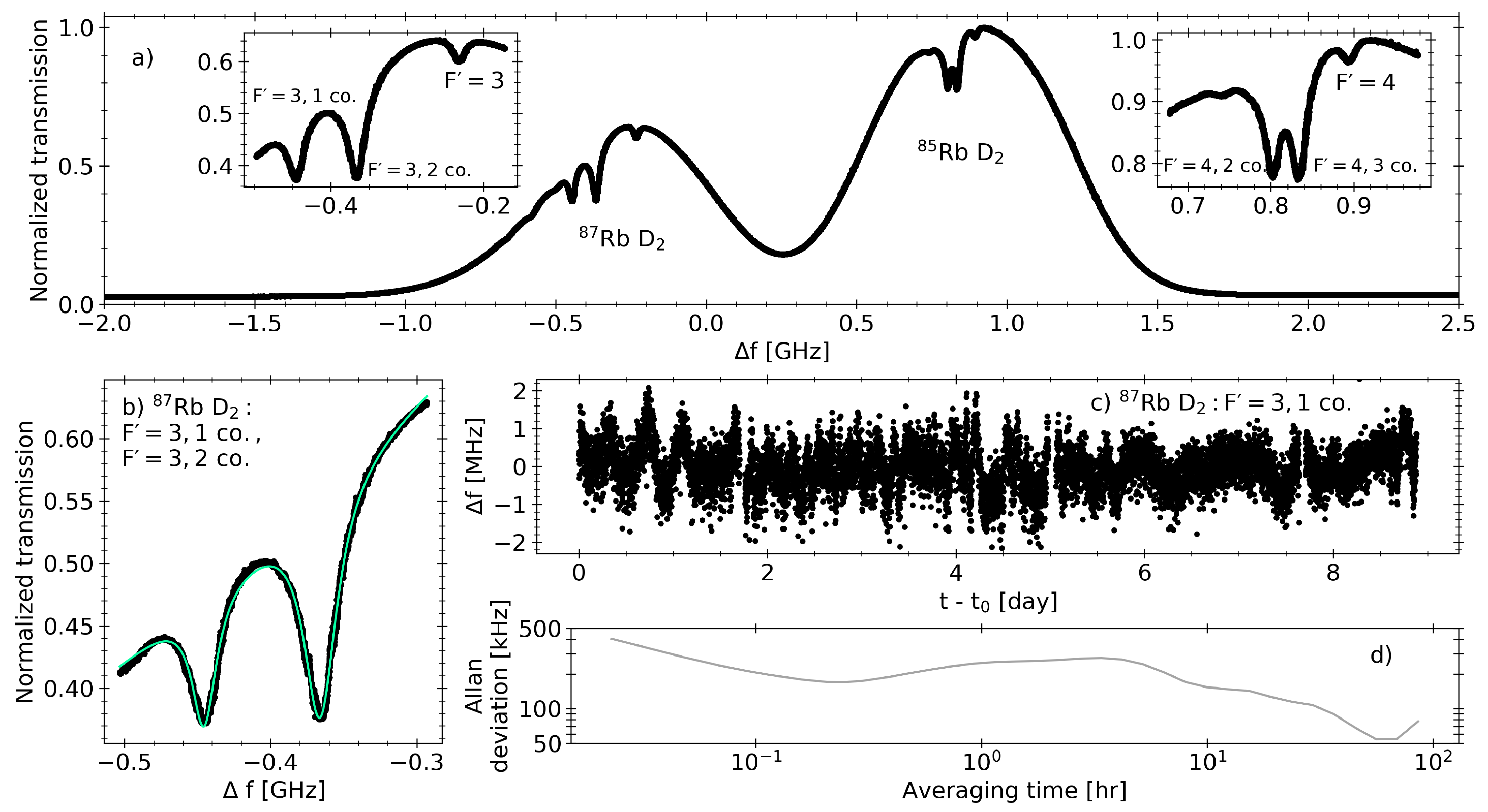}
\caption{a) Saturated absorption features of atomic rubidium lines $^{87}$Rb D$_2$ and $^{85}$Rb D$_2$ (780.24 nm) as measured over a single scan of the 780 nm laser that simultaneously traces the NIR etalon resonance. The frequency axis is arbitrarily offset by the scan centerpoint. Insets zoom on the Doppler-free hyperfine transitions (including crossover, \lq{}co.\rq{}, transitions) that superimpose on the Doppler broadened peaks. The left inset shows $F = 2 \rightarrow F'$ features for $^{87}$Rb D$_2$ and the right inset $F = 3 \rightarrow F'$ for $^{85}$Rb D$_2$. b) Fit of a sloped double Lorentzian to the $^{87}$Rb D$_2,\ F = 2 \rightarrow F' = 3,1$ co. and $F = 2 \rightarrow F' = 3,2$ co. lines. All labeled features in (a) are fit over the full timespan of Fig.~\ref{fig:drift}(a). 
c) Frequency of the $^{87}$Rb D$_2,\ F = 2 \rightarrow F' = 3,1$ co. line over the timespan of Fig.~\ref{fig:drift}(a). d) Allan deviation of the data in (c), showing no discernible linear drift down to the 50 kHz level over periods of a few days.
}
\label{fig:rubidium}
\end{figure}

Validation of our method for calibrating the frequency axes of etalon resonance scans is achieved by sending a portion of the 780 nm CW light used to track the NIR etalon resonance through a rubidium (Rb) vapor cell to simultaneously track the frequencies of several Doppler-free hyperfine transitions. Rubidium was chosen for its convenience and well known spectroscopy. {\bf Fig.~\ref{fig:rubidium}}(a) shows a single trace of the ${\rm ^{87}Rb\ D_2}$ and ${\rm ^{85}Rb\ D_2}$ lines over which the laser is scanned, and Fig.~\ref{fig:rubidium}(b) shows an example fit of two of those transitions. Fig.~\ref{fig:rubidium}(c) shows the frequency of one of these, the ${\rm ^{87}Rb\ D_2\ F' = 3,2}$ crossover transition, over the full period for which we track the 780 nm etalon resonance. In Fig.~\ref{fig:rubidium}(d) the Allan deviation of this measurement time series is decreasing out to the longest integration times, in agreement with the expectation that the vapor cell should provide a Rb spectrum stable at the level of 100 kHz. Consistent with the Allan deviation, a linear fit in Fig.~\ref{fig:rubidium}(c) shows no clear drift of the ${\rm ^{87}Rb\ D_2\ F' = 3,2}$ crossover transition greater than 10 kHz day$^{-1}$. The fluctuations seen in Fig.~\ref{fig:rubidium}(c) and (d) are consistent with the achieved signal-to-noise ratio and known systematic frequency shifts of the rubidium lines driven primarily by laser power and room temperature variations \cite{Furuta:89,AFFOLDERBACH2005,Loh:16,DINNEEN92}. This secondary frequency reference for the 780 nm light therefore independently confirms the precision of our frequency axis calibration at an order of magnitude below the etalon drift rates we measure in the NIR etalon and at least a factor of 4 below drift rates in the visible etalon (Section~\ref{sec:results}). 

\section{Results}
\label{sec:results}
While the etalons presented here are, by design, inherently simple opto-mechanical systems, the output spectrum is still sensitive to cavity temperature to some degree. We thus perform experiments under three sets of conditions. In the first, we monitor the NIR etalon resonance frequency at 1064 nm while additionally sending a broadband ($\approx 1$ nm) 976 nm laser into the cavity at optical powers of 2 mW to 20 mW to test resonance frequency response to optical heating $df/dP$ (Section~\ref{sec:sensitivity_tests}). In the second experiment, we step the temperature in the NIR etalon vacuum enclosure in $2\ {\rm^oC}$ increments to examine resonance mode thermal response $df/dT$ at 1064 nm and 1319 nm (Section~\ref{sec:sensitivity_tests}). 
In the third experiment, we monitor FP resonance frequencies under low ($<500\ {\rm \mu W}$) total input power while the cavity is under temperature control to track the drift rate at 780 nm, 1064 nm and 1319 nm in the NIR etalon (780 nm, 532 nm and 660 nm in the visible etalon). This allows us to measure the resonance frequency drift $df/dt$ as the etalon materials age over weeks (Section~\ref{sec:drift}). In all experiments we continuously record the etalon transmission peaks and heterodyne frequency marker data series.

\subsection{Optical power and temperature sensitivity}
\label{sec:sensitivity_tests}
\begin{figure}
\includegraphics[width=\columnwidth]{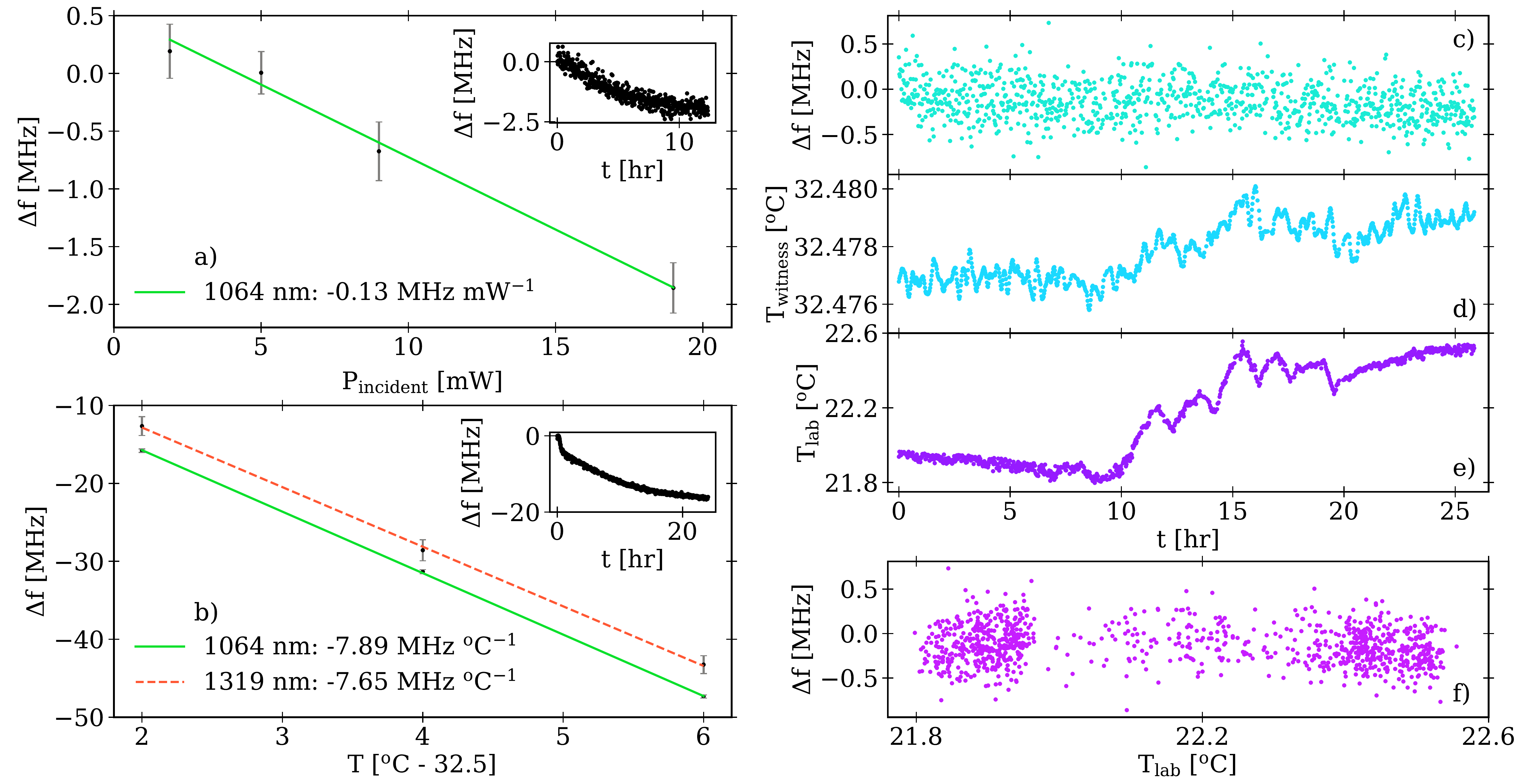}
\caption{NIR etalon. a) Frequency response of the 1064 nm resonance to heating of the cavity by a 976 nm broadband ($\approx 1$ nm) source. 
Inset: The trial at $P_{\rm incident} = 19$ mW. 
All trials have 12 hr duration. Error bars are the single measurement frequency uncertainties (Allan deviations) for each trial. b) Frequency response of the 1064 nm and 1319 nm resonances to steps in the cavity temperature setpoint. Inset: The 1064 resonance frequency response during the trial in which the temperature is stepped from 36.5 {\dC} to 38.5  {\dC}. All trials have 25 hr duration. Uncertainty on the linear fits in (a) and (b) is of order 1\%. c) -- f) Resonance frequency sensitivity to temperature in ambient conditions (when the temperature is held at $32.5\ {\rm ^oC}$ and the input power $<500\ \mu{\rm W}$). c) Relative resonance frequency at 1064 nm during a typical 25 hr segment of data. d) Temperature reported by a witness thermistor on the thermal shield surrounding the etalon. e) Ambient lab temperature. f) Resonance frequency as a function of lab temperature.}
\label{fig:sensitivities}
\end{figure}

In the first experiment we scan the 1064 nm CW across the etalon resonance at $<200\ {\rm \mu W}$ power into the cavity while additionally sending in 2 mW to 20 mW of broadband 976 nm light spanning $\approx 1$ nm. In each of four successive steps of the broadband source power, we measure the resonance frequency at fixed power to obtain a reference value, step the power, allow the cavity to equilibrate for 12 hr (we empirically find this time ensures the resonance frequency has asymptotically approached a new equilibrium value), and measure the induced shift in the resonance frequency. {\bf Fig.~\ref{fig:sensitivities}}(a) shows the resulting frequency response. A simple linear fit yields a relation for the frequency shift induced by optical heating that reaches 10 \% of the ambient frequency drift presented in Section~\ref{sec:drift} ($df/dt = 260$ kHz day$^{-1}$) at $P_{\rm incident} \approx 4.23$ mW.
Input optical powers $<1$ mW thus should not induce heating that affects measurement of the slow background drift. In trials measuring the ambient drift (Section~\ref{sec:drift}) we consequently maintain $<500\ {\rm \mu W}$ total input power so as not to appreciably affect calculated drift rates.

In the second experiment we step the temperature setpoint in the NIR cavity from 32.5 {\dC} to 38.5 {\dC} in 2 {\dC} increments. Fig.~\ref{fig:sensitivities}(b) shows the resonance response at 1064 nm and 1319 nm, with trials run for 25 hr (again an empirically determined appropriate time) to allow the resonance frequencies to equilibrate at the new temperature setpoint. These measurements yield simple estimates of the cavity CTE: $2.8 \times 10^{-8}\ {\rm ^oC^{-1}}$ and $3.4 \times 10^{-8}\ {\rm ^oC^{-1}}$ using the 1064 nm and 1319 nm data respectively. Based on the specifications of the ULE glass used for the spacer, we would expect a lower CTE in this temperature range.  However the CTE of the full cavity assembly also depends on the mirror substrates as well as the exact details of the mechanical holding structure.
These temperature step trials also indicate we are not at the CTE zero-crossing (where the cavity length response to temperature is minimized) in the NIR etalon, as the frequency response to temperature in Fig.~\ref{fig:sensitivities}(b) is roughly constant over a 6 {\dC} span.
By contrast the same experiment performed with the visible etalon shows negligible effect on the resonance frequencies, indicating this cavity is near the CTE zero-crossing.

To assess resonance mode thermal response to slower and lower amplitude ambient temperature fluctuations, Fig.~\ref{fig:sensitivities}(c) -- (f) shows the typical relationship between frequency of the 1064 nm mode and the ambient temperature when the etalon temperature is held constant. This 25 hr segment shows a $\approx 5\ {\rm mK\ ^o C^{-1}}$ response of the witness thermistor to lab temperature fluctuations. 
Yet Fig.~\ref{fig:sensitivities}(f) demonstrates that this weak tracking of the witness temperature with lab temperature causes no discernible systematic effect on the resonance frequency.
This is consistent with what one would expect from the theoretical CTE, where a 5 mK change would correspond to a $\approx 50$ kHz resonance frequency shift.
We thus do not find thermal sensitivity to variations in lab temperature to substantially affect our measured drift rates.

\subsection{Frequency drift}
\label{sec:drift}
\begin{figure}
\includegraphics[width=\columnwidth]{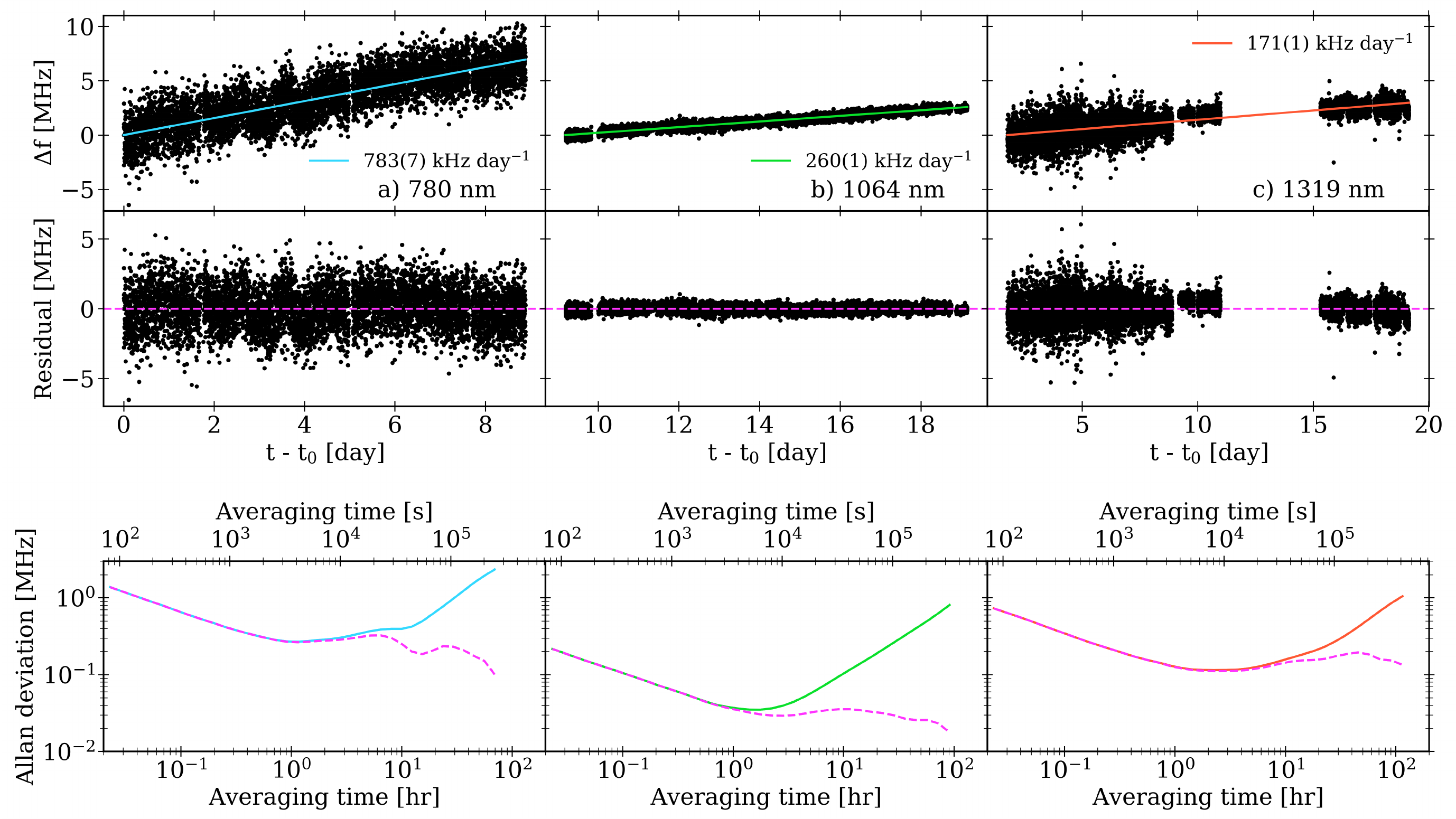}
\caption{NIR etalon. a) Frequency drift $df/dt$ of the etalon resonance at 780 nm as the materials age, with a linear fit shown whose slope and uncertainty as reported by the linear regression are given in the legend. Residuals to this fit are shown below, as well as Allan deviations for the frequency stability of the resonance frequency with the drift (solid line) and for the residuals (dashed). b -- c) As in (a) for the 1064 nm and 1319 nm resonances. Drift rates are examined in Section~\ref{sec:drift}. The reduced scatter in the 1319 nm data after day 9 is a result of improved SNR. In the NIR setup, the 1319 nm resonance is scanned during all measurement periods, while either the 780 nm or 1064 nm resonance is selected for a specific measurement.}
\label{fig:drift}
\end{figure}

The frequency (or wavelength) stability of the FP etalon modes over time is the most important characteristic that our measurements allow us to address. During these measurements the temperature of the etalon thermal shield is being actively controlled at its fixed point, but otherwise the outer vacuum housing experiences the ambient laboratory environment, with the lab temperature fluctuating over a maximum range of $\approx 1$ {\dC}. Using the techniques described in Section~\ref{sec:lab_charac_scheme} we measure the drift of the NIR resonance frequencies at 780 nm, 1064 nm and 1319 nm over periods $>1$ week.  Results are shown in {\bf Fig.~\ref{fig:drift}}, with a linear fit to the measured mode frequencies over time yielding observed drift rates of $df/dt =$783(7) kHz day$^{-1}$, 260(1) kHz day$^{-1}$ and 171(1) kHz day$^{-1}$ (see {\bf Table~\ref{tab:drift_rates}} for a summary of these and the following quantities). These correspond to a fractional frequency stability of the respective etalon modes of $2 \times 10^{-9}$, $9 \times 10^{-10}$ and $8 \times 10^{-10}$ day$^{-1}$. A simple model would predict that the ratio of the drift rates should scale as the ratio of their mode numbers ($df_{780} / df_{1064} = 1.37$, $df_{1064} / df_{1319} = 1.24$ and $df_{780} / df_{1319} = 1.70$; Section~\ref{sec:background}). However the observed ratios differ markedly from this at 3.02(0.03), 1.52(0.01) and 4.57(0.05) respectively. This discrepancy cannot be explained by uncertainties in our measurement approach, and possible physical mechanisms are discussed below. After removing a linear fit to the drift we calculate the Allan deviation, a measure of the frequency stability as a function of averaging time. As shown in Fig.~\ref{fig:drift}, this indicates that under the assumption of linear drift we are able to determine the line center at all sample resonances to better than 300 kHz for averaging periods of 1 hr up to the maximum averaging time of $\approx 100$ hr. Importantly the different scatter in the resonance measurements and the computed Allan deviations are indicative of the measurement SNR rather than an intrinsic property of a specific resonance.

\begin{table*}
\caption{Theoretical and measured drifts of the near infrared and visible etalons under investigation. See Section~\ref{sec:drift} for an explanation and Section~\ref{sec:discussion} for a discussion of these values.}
\begin{center}
\begin{tabular}{lcc}
\hline
& \bf{NIR etalon} (HPF) & \bf{Visible etalon} (NEID)  \\
\hline
\multirow{3}{7em}{Measured fractional frequency stability [day$^{-1}$]} & 780 nm: $2 \times 10^{-9}$ & 780 nm: $9 \times 10^{-11}$\\
& 1064 nm: $9 \times 10^{-10}$ & 532 nm: $2 \times 10^{-10}$\\
& 1319 nm: $8 \times 10^{-10}$ & 660 nm: $1 \times 10^{-10}$\\
\multirow{3}{7em}{Measured drift rate [kHz day$^{-1}$]} & 780 nm: 783(7) & 780 nm: 36(1) \\ 
& 1064 nm: 260(1) & 532 nm: 129(2) \\ 
& 1319 nm: 171(1) & 660 nm: 48(2) \\
\multirow{3}{7em}{Theoretical drift ratio} & $df_{780} / df_{1064} = 1.37$ & $df_{532} / df_{780} = 1.47$ \\ 
& $df_{1064} / df_{1319} = 1.24$ & $df_{532} / df_{660} = 1.24$ \\ 
& $df_{780} / df_{1319} = 1.70$ & $df_{660} / df_{780} = 1.18$ \\
\multirow{3}{7em}{Measured drift ratio} & $df_{780} / df_{1064} = 3.02(0.03)$ & $df_{532} / df_{780} = 3.60(0.11)$ \\ 
& $df_{1064} / df_{1319} = 1.52(0.01)$ & $df_{532} / df_{660} = 2.66(0.12)$ \\ 
& $df_{780} / df_{1319} = 4.57(0.05)$ & $df_{660} / df_{780} = 1.36(0.07)$ \\
\hline
\end{tabular}
\end{center}
\label{tab:drift_rates}
\end{table*}

\begin{figure}
\includegraphics[width=\columnwidth]{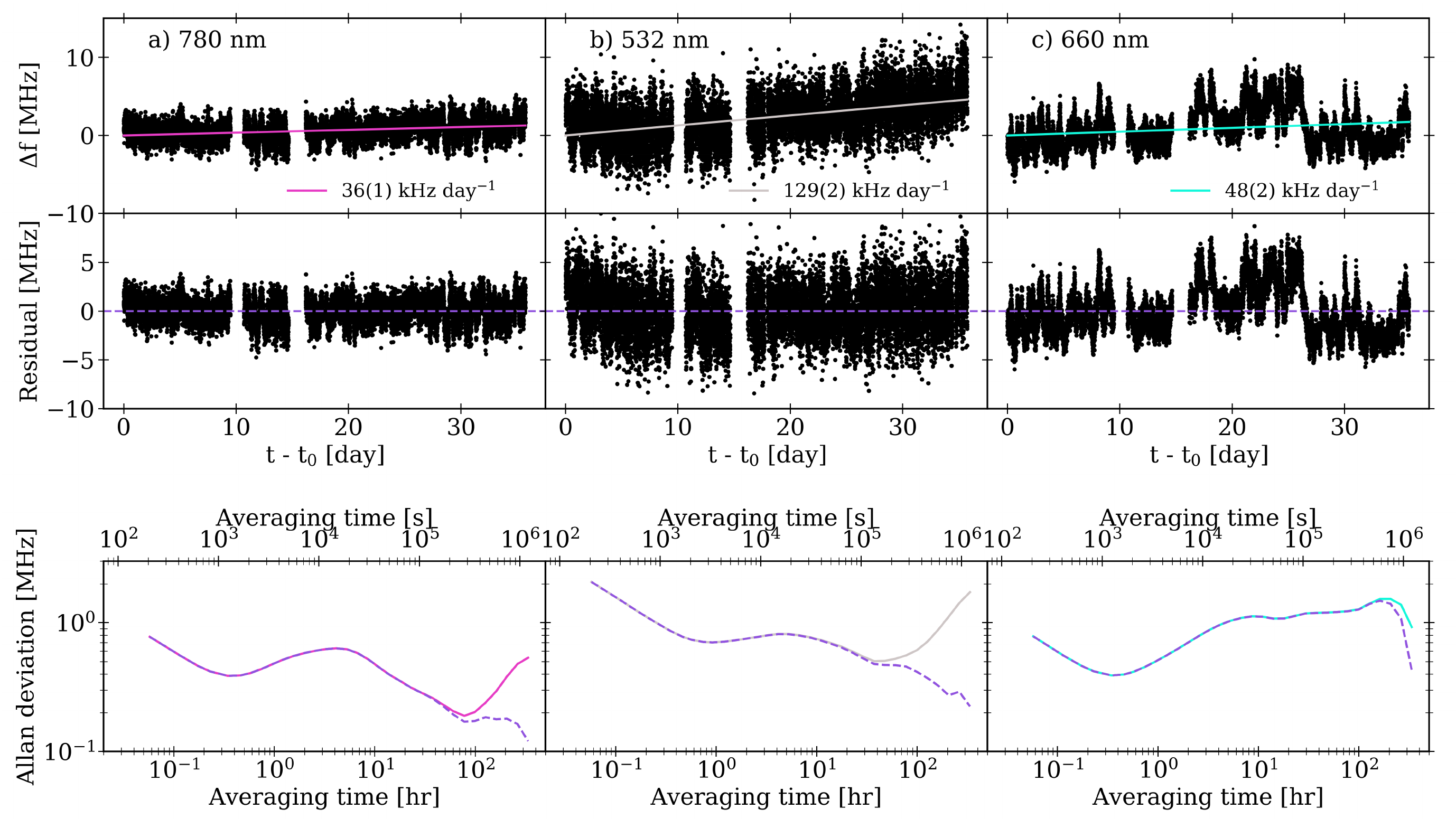}
\caption{Visible etalon. a) Frequency drift $df/dt$ of the etalon resonance at 780 nm as the materials age, with a linear fit shown whose slope and uncertainty as reported by the linear regression are given in the legend. Residuals to this fit are shown below, as well as Allan deviations for the frequency stability of the resonance frequency with the drift (solid line) and for the residuals (dashed). b -- c) As in (a) for the 532 nm and 660 nm resonances. 
}
\label{fig:drift_visible}
\end{figure}

Analogously {\bf Fig.~\ref{fig:drift_visible}} shows the ambient drift in the visible etalon at 780 nm, 532 nm and 660 nm. Here the respective observed drift rates, $df/dt =$36(1) kHz day$^{-1}$, 129(2) kHz day$^{-1}$ and 48(2) kHz day$^{-1}$, correspond to a frequency stability of the respective etalon modes of $9 \times 10^{-11}$ day$^{-1}$, $2 \times 10^{-10}$ day$^{-1}$ and $1 \times 10^{-10}$ day$^{-1}$ (see Table~\ref{tab:drift_rates} for a summary). Again a simple model of the FP would predict drift rate ratios of $df_{532} / df_{780} = 1.47$, $df_{532} / df_{660} = 1.24$ and $df_{660} / df_{780} = 1.18$, but instead we observe 3.60(0.11), 2.66(0.12) and 1.36(0.07) respectively. The larger Allan deviations at a given averaging time relative to those in the NIR etalon data are due to lower SNR in the visible etalon measurement setup. In particular the measurements of the 660 nm resonance data show increased fluctuations when compared to the measurements at other wavelengths, likely due to increased sensitivity to laser amplitude noise and laboratory temperature perturbations. 

\begin{figure}
\includegraphics[width=\columnwidth]{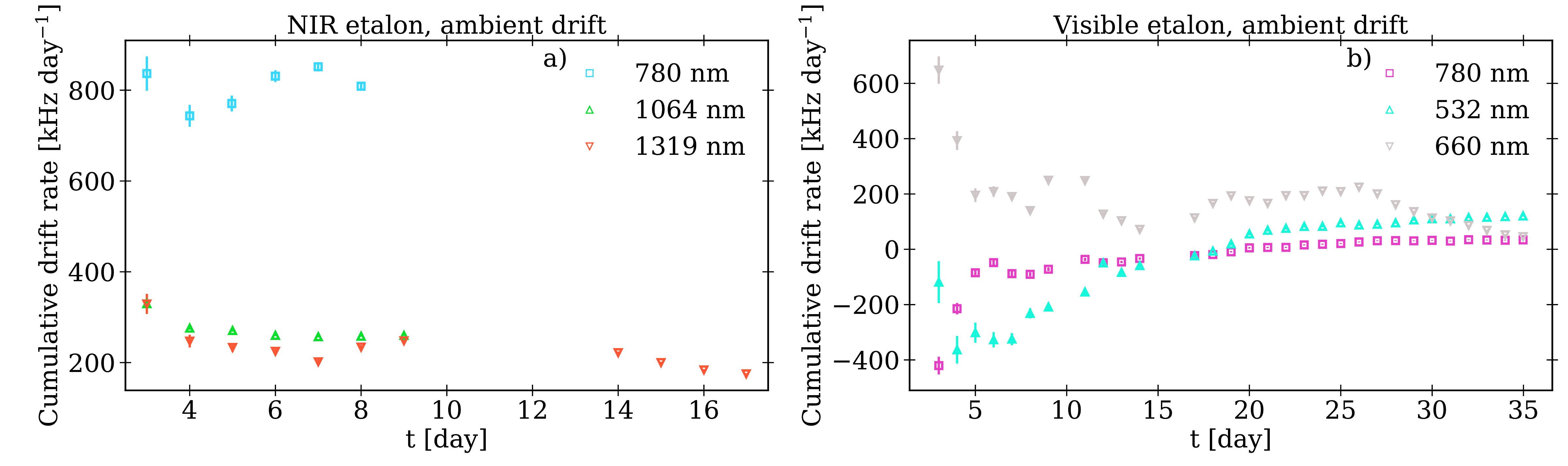}
\caption{NIR and visible etalons. a) Frequency drift rate $df/dt$ for the NIR etalon resonances at 780 nm, 1064 nm and 1319 nm, shown in successively larger cumulative portions of the full dataset to demonstrate how the rates evolve as observation time increases. All points have $1\sigma$ error bars. In the NIR setup, the 1319 nm resonance is scanned during all measurement periods, while either the 780 nm or 1064 nm resonance is selected for a specific measurement. b) As in (a) for the visible etalon resonances at 780 nm, 532 nm and 660 nm.}
\label{fig:drift_evo}
\end{figure}

{\it The ability to accurately predict the drift rate of the FP cavity resonances is critical for their use as standalone frequency calibrators.} {\bf Fig.~\ref{fig:drift_evo}}(a)--(b) show how the measured drift rates in both etalons evolve as we fit a straight line for the linear drift to successively larger portions of the cumulative datastream. As would be expected, the ability to predict the drift rates improves with measurement time. And while the data for the NIR cavity is limited in duration, the visible etalon data shows evidence of a nonlinear change in the drift rate, at least over the measurement time of $\approx 1$ month. We note that measurements with the visible etalon were begun within a few weeks of the mirrors being optically contacted. This slowing of the drift rate could be indicative of the cavity "relaxing" in its mechanical structure and/or a slow change in the cavity length itself. Particularly encouraging for the visible cavity is that the drift rates at all three colors are evolving to values $\lesssim 100$ kHz day$^{-1}$, equivalent to fractional drift rates $\lesssim 2\times 10^{-10}$ day$^{-1}$. For astronomical RV calibrations this corresponds to 5 cm s$^{-1}$ day$^{-1}$.

\section{Discussion}
\label{sec:discussion}
The experiments and data reported above demonstrate the excellent frequency stability that can be obtained from broad bandwidth FP etalons designed for astronomical spectrograph calibration. Slow, predictable and largely linear drifts should enable such FP etalons to serve as efficient standalone calibrators on timescales of days -- weeks, and cross calibration with other sources (lamps, gas cells or frequency combs) should provide enhanced long-term stability. However the most striking observation of our work is the discrepancy between the theoretical and observed drift ratios, i.e., an observed chromatic variation in the linear drift that does not fit the simple model of a FP etalon. Our measurements indicate a wavelength dependence of the FP modes beyond that predicted by the ratio of mode numbers, which cannot be accounted for by the uncertainties in the present laboratory measurements. 
The significance of our current result is that {\it knowledge (and control) of the frequency of a single etalon mode does not reliably predict the frequencies of other modes, especially over broad bandwidths.} This observed behavior could arise from one or a combination of multiple aspects of the etalon's construction and illumination. While we do not have conclusive results on the exact mechanism, our experiments do allow us to suggest some possibilities. 

One of the most striking properties of the plane-parallel FP cavity is its sensitivity to input alignment -- that is, the degree of parallelism between the input beam and the cavity's optical axis. In experiments conducted before the visible etalon was placed in vacuum, the CW lasers were coupled into a single mode fiber, the output of which was collimated and aligned to the cavity by maximizing the power reflected back into the same optical fiber. We then scanned the incidence angle onto the FP in one dimension by adjusting a kinematic mirror mount before the etalon and sampled the resulting resonance frequency shifts at our three colors. In this experiment we measured $\sim 1$ MHz $\mu$rad$^{-1}$ sensitivity of the mode frequencies at 780 nm, 660 nm and 532 nm to input alignment. 
This suggests that wavelength-dependent alignment errors $\leq 1\ \mu$rad varying differently in time could result in frequency shifts with magnitude comparable to that observed in our ambient drift measurements. For example, although care was taken in collimation of the broad bandwidth light incident on the cavity, small differences in angular divergence could lead to different colors sampling different spatial regions of the cavity. This could then couple to spatial variations in the surface figure of the mirrors or to a relaxing of the cavity that is not radially or azimuthally symmetric relative to the cavity axis. Such a hypothesis is difficult to model or verify experimentally, but it would be informative if revealed in other FP cavities of similar construction that are in use or intended for similar application. We note that in a previous experiment \cite{2017OExpr..2515599J} we did recover the expected dispersion stability $\Delta \omega_{1064} / \Delta \omega_{1319}$ at the $10^{-3}$ level in a fiber {\fp} using the measurement technique and data analysis pipeline we employ here. Though note for the fiber {\fp}, the fiber is single mode at all wavelengths, thus ensuring ideal spatial overlap of all wavelengths.

Another possible cause of the observed anomalous wavelength dependence of the etalon drift is the frequency-dependent phase upon reflection. For example if the reflective phase shift is strongly varying near one or more of the three test wavelengths we measure, then as the cavity length changes by $\Delta L$ the different resonant frequencies would be given by 
\begin{equation} 
   \frac{\Delta \omega_m}{\Delta L}  = - \frac{c} {L^2}(\pi m - \phi_{r}) - \frac{c}{L}\frac{\Delta \phi_r}{\Delta L},
 \label{eqn:deriv}
 \end{equation}
which follows by differentiation of Eq.~\ref{eqn:omega}. From the theoretical reflective phase shift for the NIR cavity mirrors, we find that $\phi_r$ varies by $\approx 10\pi$ across the 800 -- 1300 nm reflectivity band. This is however much smaller than the first term in the parentheses of Eq.~\ref{eqn:deriv}, which is on the order of $10^4 \pi$. Thus the direct impact of $\phi_r$ does not appear to explain our observations. Eq.~\ref{eqn:deriv} also shows that there could be an additional effect arising from $\Delta \phi_r/\Delta L$.  Although it seems unlikely that $\phi_r$ depends on $\Delta L$, unless there is some stress-induced coupling between the etalon spacer length and the reflective phase shift.  
At present the phase response of the mirror coatings to temperature and mechanical stress (or relaxation) are unknown. The phase shift may even be time varying as a result of the mirrors being placed in vacuum, as posited by \cite{Riehle98}. 
The impact of the mirror structure itself on the cavity drift thus remains a topic of interest that merits further investigation. Future data and analysis on the long-term wavelength dependent drift from both etalons at their respective telescopes, where their full spectra are being measured with high precision spectrographs calibrated with LFCs, may help discern what is causing the anomalous wavelength dependence of the etalon drift. In particular, while less direct than our measurement technique, this approach will have different systematic uncertainties that may allow us to exclude one or more of the potential causes discussed above. 
 
\section{Conclusions}
We have described the construction and characterization of two evacuated, broadband, planar {\fp} etalons that are to be used as spectral calibrators for precision astronomical RV spectroscopy. Using an auxiliary laser frequency comb, we track three resonances in each etalon over a time frame of 1 week to 1 month. We observe that the drift rates of the cavity resonances are slow and essentially linear over this same time frame, with magnitudes $< 2\times 10^{-9}$ day$^{-1}$.  However we also find that the ratios of drift rates among the monitored resonances differ significantly from the theoretical expectation. 
The cause is likely a combination of factors including alignment sensitivity and cavity imperfections over extremely broad bandwidths, as well as unknown factors related to the details of the dispersion and stability of the cavity mirror coatings. An important outcome of our work is that tracking one resonance mode of such a planar etalon is likely insufficient to provide full information about the frequency drift of other modes via a simple ratio of the frequencies. This may be of importance to the range of users in the astronomical community who have adopted etalon-based calibrators.

\section*{Funding}
We acknowledge support from NSF grants AST-1006676, AST-1126413, AST-1310875, AST-1310885, AST-1517592, as well as NIST. NEID is funded by NASA through JPL by contract 1547612.

\section*{Acknowledgments}
The mention of trade names and specific products is for scientific information only and does not imply an endorsement of such products by NIST. This work was performed by SPH [in part] under contract with the Jet Propulsion Laboratory (JPL) funded by NASA through the Sagan Fellowship Program executed by the NASA Exoplanet Science Institute. We thank Rich Fox and Daniel Nicolodi for their thoughtful comments on the manuscript.  JJ thanks A. Bagga for his discussions on the work.

\section*{Disclosures}
MG, MN: Stable Laser Systems (E).

\appendix
\section{Etalon resonance residual structure}
\label{sec:residual_structure}

\begin{figure*}
\includegraphics[width=\columnwidth]{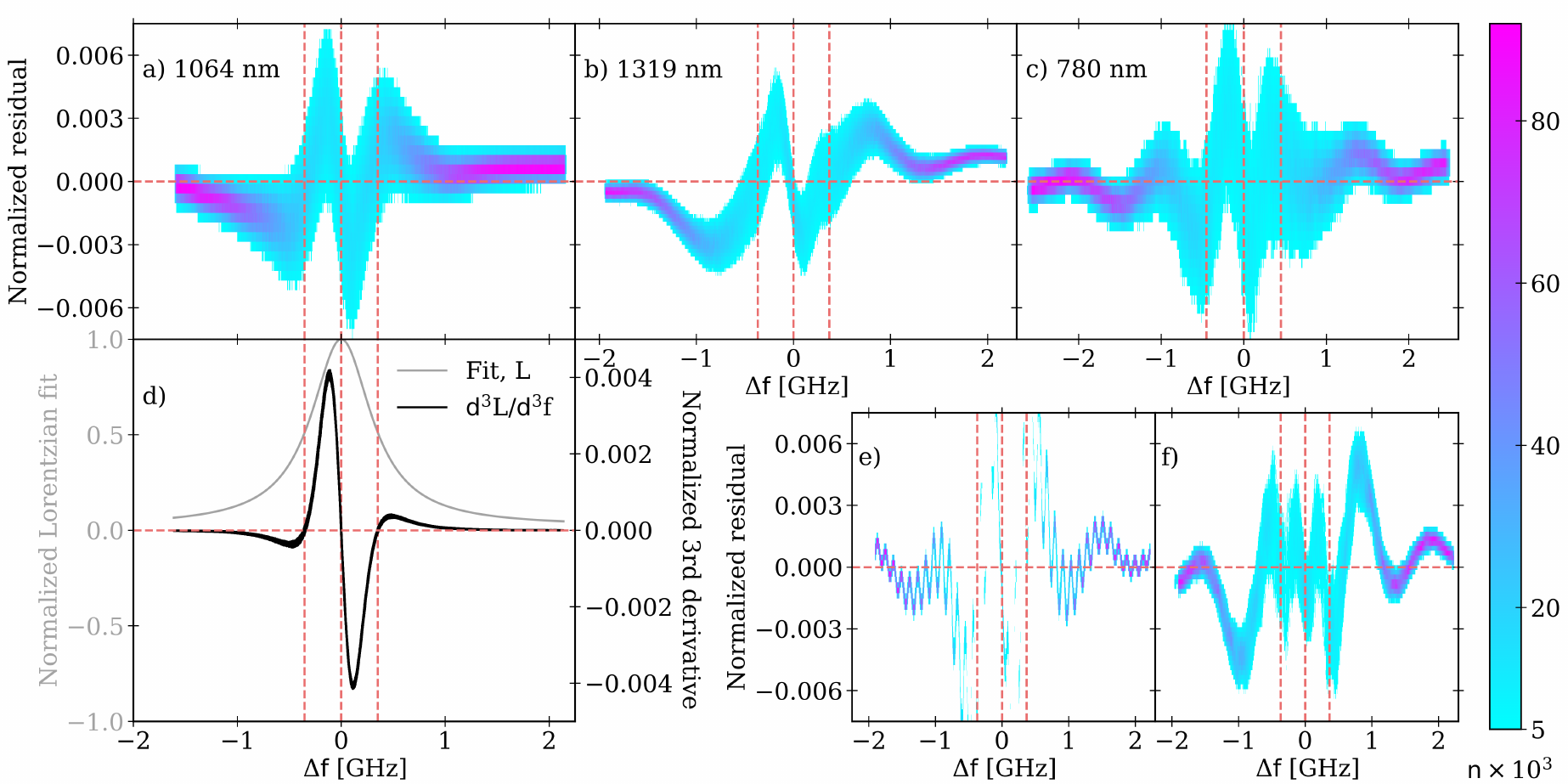}
\caption{NIR etalon. a) 2D histogram of residuals to a Lorentzian fit on the 1064 nm resonance for 12 hr of continuous data ($\approx 500$ resonance scans; compare with residuals for a single scan in Fig.~\ref{fig:resonances}(b)). Residuals are normalized to the mean Lorentzian fit amplitude across all scans. Vertical lines show the mean fit linecenter and linewidth. Data are binned in 10 MHz intervals and 100 bins in normalized amplitude. The linear colorbar at right (common to all subplots excluding (d)) shows counts in each bin, with a minimum of 5000 counts. b) As in (a) for the 1319 nm resonance (compare with residuals for a single scan in Fig.~\ref{fig:resonances}(c)). c) As in (a) for the 780 nm resonance (compare with residuals for a single scan in Fig.~\ref{fig:resonances}(a)). d) A Lorentzian profile fit to one of the resonance scans used to produce (a), as well as the fit's third derivative. The third derivative is qualitatively similar to the structure in (a), as discussed in Appendix A. e) Parasitic etaloning can alter the structure in (a) -- (c). Here is shown a case of parasitic etaloning at 1319 nm, with 3 hr of continuous data used to produce the histogram. This demonstrates both the parasite stability on the same timescale and its ability to set the envelope of the residuals in (a) -- (c).  f) A case of more severe parasitic etaloning at 1319 nm (using 12 hr of continuous data), showing the distortion to the underlying residual structure in (b).}
\label{fig:residual_struc}
\end{figure*}

Residuals to a Lorentzian fit of the etalon resonances in both the NIR and visible cavities show stable structure of unknown origin. In {\bf Fig.~\ref{fig:residual_struc}}(a) -- (c), 2D histograms show this systematic structure at the $<1\%$ level over 12 hr ($\approx 500$ stacked 40 s resonance scans) in the NIR etalon resonance fit residuals at 1064 nm, 1319 nm and 780 nm respectively. 
We verified that the structure is not due to approximating the resonance Airy profile by a Lorentzian; for the data in Fig.~\ref{fig:residual_struc}, the difference of an Airy fit and our Lorentzian fit at each of the 3 test wavelengths does not exceed $10^{-5}$ fractional of the resonance profile's amplitude (i.e., is a factor of 1000 below the systematic residual structure). The stable residual structure is instead likely a consequence of imperfections that distort the transmission profile, such as mirror surface defects that would cause a position-dependent mirror separation \cite{1963AcOpt..10..141H, 1985ApOpt..24.1502W, 2008A&A...481..897R} or slight mirror non-parallelism that would produce an asymmetric line profile \cite{1963AcOpt..10..141H, 1986ApOpt..25.1646M}. It is unclear whether this structure is a consequence of the same mechanism(s) causing the anomalous wavelength dependence of the etalon drift. Fig.~\ref{fig:residual_struc}(d) shows the similarity of the residual structure to the third derivative of the Lorentzian fit to a single resonance scan at 1064 nm. Parasitic etalons are also seen distorting the stable residual structure in Fig.~\ref{fig:residual_struc}(e) -- (f). In many cases these could be removed by adjusting the upstream optics. For those that could not be removed, we test their efficacy to bias our resonance fits and explain the anomalous wavelength dependence of the etalon drift. For the data in Fig.~\ref{fig:residual_struc}(a) for example, we take the residual amplitude with the largest count at each bin along the frequency axis and subtract this from the raw resonance profile in each of the $\approx 500$ scans used to produce the plot. For each scan we then fit a Lorentzian to this data from which the stable residual structure is subtracted and compare the fitted linecenter with  that obtained by fitting to the raw data. The difference between the fitted linecenters is $<1$ kHz, demonstrating that this stable residual structure does not systematically affect our measured drift rates.

\bibliography{refs}

\end{document}